\documentstyle[amstex,amssymb,preprint,aps]{revtex}
\begin{document}
\tightenlines
\draft
\title{On Some Exponential Potentials for a Cosmological Scalar Field as
Quintessence}
\author{Claudio Rubano}
\address{Dipartimento di Scienze Fisiche, Universit\`{a} Federico II\\
and \\
Istituto Nazionale di Fisica Nucleare, Sez. di Napoli,\\
Complesso Universitario di Monte S. Angelo,\\
Via Cintia, Ed. G, I-80126 Napoli, Italy\\
}
\author{Paolo Scudellaro}
\address{Dipartimento di Scienze Fisiche, Universit\`{a} Federico II\\
and \\
Istituto Nazionale di Fisica Nucleare, Sez. di Napoli,\\
Complesso Universitario di Monte S. Angelo,\\
Via Cintia, Ed. G, I-80126 Napoli, Italy}
\date{\today}
\maketitle

\begin{abstract}
We present general exact solutions for two classes of exponential
potentials in a scalar field model for quintessence. The coupling is
minimal and we consider only dust and scalar field. To some extent, it is
possible to reproduce experimental results from supernovae.

PACS number(s): 98.80.Cq, 98.80.Hw, 04.20.Jb

\end{abstract}

\narrowtext
\section{Introduction}

In the last few years, new models of the universe have been built taking
{\em dark energy} into account \cite{van,Turner1}. Together with baryons,
cold dark matter, photons and neutrinos, a fifth component has been added,
the socalled {\it quintessence} field $Q$ \cite{Ostrik,CDS,ZWS,SWZ,Wang}
(or, in general, the {\em x-field } \cite{Turner2,Chiba}). Generalizing
ideas like that of a cosmic equation of state variable with a
$\Lambda$-term \cite{Zel}, with respect to a more usual cosmological
constant $\Lambda$, such a $Q$-field, even if still implying a negative
pressure contribution to the total pressure of the cosmic fluid, is
characterized by the fact that its equation of state is given by
$-1<w_{Q}\equiv p_{Q}/\rho_{Q}<0$, $p_{Q}$ and $\rho _{Q}$ being,
respectively, the pressure and energy density of the $Q$-field. Actually,
the interval $-1<w_{Q} \lesssim -0.6$ is usually considered \cite{Perl1}.
As a matter of fact, when $w_{Q}=-1$ we recover a constant $\Lambda $-term
\cite{Carroll,Sahni1,Bin}, which can be regarded as a measure of vacuum
energy density, leading to the well known discrepancy between theory and
observations \cite{Wein,Carroll,Sahni1}, based on the question of why
$\rho_{Q}$ is so small with respect to typical particle physics scales.
(But there are also mechanisms of relaxation of the cosmological constant
during the initial inflationary stage, which could explain such a
discrepancy; see \cite{Rubakov}, for instance.)

As well known, an interesting possibility to handle the presence of
quintessence in the universe is to see it as given by a scalar field
$\varphi $ slowly rolling down its potential $V(\varphi )$. If we define
\begin{equation}
\rho _{\varphi }\equiv \frac{1}{2}\dot{\varphi}^{2}+V(\varphi )\quad ;\quad
p_{\varphi }\equiv \frac{1}{2}\dot{\varphi}^{2}-V(\varphi ),  \label{eq0}
\end{equation}
(dot indicating time derivative, and $V(\varphi )$ being the potential for
$\varphi $), the slow rolling condition immediately gives $p_{\varphi }<0$
and $w_{\varphi }\equiv p_{\varphi }/\rho _{\varphi }\simeq -1$. With such
a negative pressure, the universe evolves like in a sort of present day
{\em soft} inflationary scenario, so allowing to explain observations on
supernovae \cite{Perl2,Perl3,Riess,Garn,Jha} and why vacuum and matter
densities are today comparable (`cosmic coincidence' problem
\cite{Stein,Carroll,Sahni1}). (Alternatively to quintessence, a negative
pressure and an explanation of current observations can also be obtained
in a {\em Chaplygin cosmology} \cite{Kamen}.)

Many cosmological models with a dynamical scalar field have been proposed,
showing {\em scaling} solutions, i.e., such that at some time $\rho _{m}$
and $\rho _{\varphi }$ simultaneously depend on some powers $n_{1}$ and
$n_{2}$ of the scale factor $a$, acting as attractors in the phase space.
When $n_{1}=n_{2}$, we have the socalled {\em self-tuning} solutions
\cite{Liddle}, which are typically driven by exponential potentials. This
kind of potential has been studied extensively
\cite{Ratra,Peebles,Wett1,Wett2,Cope,Ferr,Fabris,Barr,Sahni2,Brax},
especially from a qualitative point of view (see \cite{Batista} and
references therein, for instance).

The simplest possibility is of course $V(\varphi )=\alpha
e^{\lambda\varphi}$, which is often discarded (see discussion in
\cite{SWZ,Bin}). More promising seems to be a combination of two terms
$V(\varphi )=\alpha e^{\lambda \varphi }+\beta e^{-\lambda \varphi }$
\cite{Barr}.

In this paper, we consider a particular class of both these types from a
different point of view, obtaining {\em general exact solutions}. This
allows a very stringent comparison with experimental data on supernovae, so
that also the first type seems to deserve further investigations; for the
second type, we obtain a solution which can mimic very well the presence of
a cosmological constant in the late evolution of the universe. Both of them
are not, strictly speaking, {\it scaling solutions}, although this concept
may be recovered in a more general sense.

Another experimental fact which we use is the strong evidence of a
spatially flat universe \cite{Nature}. Thus, we set the scalar curvature
$k=0$ in all our equations from the very beginning. However, as we shall
see, the values of $\Omega _{m}$ and $\Omega _{\Lambda }$ ($\Omega\varphi$
in our case) derived from the experiments strongly depend on the model, so
that some discussion is needed.

Mostly, the scalar field $\varphi$ has been considered as minimally
coupled to gravity, even if (more recently) a nonminimal coupling has also
been introduced \cite{Uzan,Amendola,Chiba1,PBM,Bertolami,deR1}. Here, we
will consider a very simple model consisting of a two-component
cosmological fluid: matter and scalar field. `Matter' means baryonic +
cold dark matter, with no pressure, and the scalar field is minimally
coupled and noninteracting with matter. Clearly, this model cannot be used
from the very beginning of the universe, but only since decoupling of
radiation and dust. Thus, we do not take into account inflation, creation
of matter, nucleosynthesis, etc. The main shortcut is that we cannot
really check for the {\em tracker} feature \cite{ZWS,SWZ} of the
$\varphi$-solution.

In Sec. II we take a particular exponential potential into account,
deriving the general exact solution of the cosmological equations and,
thereby, cosmological parameters as functions of time, so allowing the
comparison with observational data. Sec. III is devoted to the same kind
of considerations for a potential given by a linear combination of two
such exponentials. Conclusions are drawn in Sec. IV.

\section{An exponential potential}

\subsection{Mathematical treatment}

Let us consider a spatially flat, homogeneous and isotropic universe,
filled with two noninteracting components only, i.e., pressureless matter
(or {\em dust}) and a scalar field $\varphi$, minimally coupled to
gravity. The cosmological equations are then
\begin{gather}
3H^{2}=\frac{8\pi G}{c^{2}}(\rho _{m}+\rho _{\varphi }),  \label{eq1} \\
\dot{H}+H^{2}=-\frac{4\pi G}{3c^{2}}(\rho _{m}+\rho _{\varphi
}+3(p_{m}+p_{\varphi })),  \label{eq2} \\
\ddot{\varphi}+3H\dot{\varphi}+V^{\prime }(\varphi )=0,  \label{eq3}
\end{gather}
where prime indicates derivative with respect to $\varphi$, $H \equiv
\dot{a }/a$ is the Hubble parameter, $p_{m}=w_{m}\rho _{m}$ and
$p_{\varphi }=w_{\varphi }\rho _{\varphi }$ are the equations of state for
matter and scalar field. Let us stress that $w_{\varphi }$ is {\em \ not
constant}, and that we want to describe some features of cosmology after
the decoupling.

We set $w_{m}=0$, so that $\rho _{m}=Da^{-3}$. The parameter $D \equiv
{\rho_{m}}_{0}{a_{0}}^{3}$ (the lower index `${}0$' indicating present day
values) is the amount of matter. The equations can also be rewritten as
\begin{gather}
\left( \frac{\dot{a}}{a}\right) ^{2}=\frac{8\pi G}{3c^{2}}(Da^{-3}+\frac{1}{2
}\dot{\varphi}^{2}+V(\varphi )),  \label{eq1a} \\
2\frac{\ddot{a}}{a}+\left( \frac{\dot{a}}{a}\right) ^{2}=-\frac{8\pi
G}{3c^{2}}\left( \frac{1}{2}\dot{\varphi}^{2}-V(\varphi )\right) ,
\label{eq2a}
\\
\ddot{\varphi}+3\left( \frac{\dot{a}}{a}\right) \dot{\varphi}+V^{\prime
}(\varphi )=0.  \label{eq3a}
\end{gather}

In this Section, we consider the potential
\begin{equation}
V(\varphi )=B^{2}e^{-\sigma \varphi },  \label{eq4}
\end{equation}
where $B^{2}$ is a generic positive constant and
\begin{equation}
\sigma ^{2}\equiv \frac{12\pi G}{c^{2}}.  \label{eq5}
\end{equation}
(The minus sign in the exponential is irrelevant, since there is symmetry
with respect to a change $\varphi \rightarrow -\varphi$.)

This type of potential leads to a late time attractor in a scalar-field
dominated situation ($\Omega _{\varphi }=1,w_{\varphi }=-0.5$)
\cite{Bin,Barr}. Being aware of such a behaviour, anyway, we stress that
we are especially interested in the contemporary or, at most, the recent
past regimes, where the situation is different. Usually, associated with
an exponential potential, a scalar field is considered such that
$\Omega_{\varphi }\equiv 8\pi G\rho _{\varphi }/(3c^{2}H^{2})$ is
practically constant during part of the matter-dominated era. This implies
that assuming $w_{\varphi }\sim constant$ leads to a constant ratio of
quintessence to matter energy density, so that $\Omega _{\varphi }$ (being
$\lesssim 0.15$ at the beginning of matter-dominated era, due to
nucleosynthesis \cite{Cope,Ferr}) must remain small forever
\cite{SWZ,Bin}. Mainly for such reasons, this kind of potential is not
considered as suitable for a quintessence field.

The particular choice of Eq. (\ref{eq5}) for $\sigma $ allows for general
exact integration of equations. Such a choice was in fact used in the
context of inflationary theory by us \cite{deR2,RivNC} and others
\cite{Barrow1,Barrow2}, with a scalar field only.

Let us concentrate on the second order equations (\ref{eq2a}) and
(\ref{eq3a}), while Eq. (\ref{eq1a}), which is a first integral, is
considered as a constraint on the integration constants. Let us introduce
the new variables $u$ and $v$, defined by the tranformation
\begin{equation}
a^{3}=uv\quad ;\quad \varphi =-\frac{1}{\sigma }\log \frac{u}{v},
\label{eq6}
\end{equation}
which is always invertible (the Jacobian being $J=2/\sigma $). We get for
the potential
\begin{equation}
V(u,v)=B^{2}\frac{u}{v},
\end{equation}
and Eqs. (\ref{eq2a}) and ( \ref{eq3a}) become
\begin{equation}
\ddot{u}=0\quad ;\quad \ddot{v}=\omega u,  \label{eq7}
\end{equation}
where $\omega =\sigma ^{2}B^{2}=12\pi GB^{2}/c^{2}>0$. They are immediately
integrated to
\begin{eqnarray}
u(t) &=&u_{1}t+u_{2},  \label{eq8} \\
v(t) &=&\frac{1}{6}u_{1}\omega t^{3}+\frac{1}{2}u_{2}\omega
t^{2}+v_{1}t+v_{2},  \label{eq9}
\end{eqnarray}
being $u_{1}$, $u_{2}$, $v_{1}$, and $v_{2}$ arbitrary integration
constants. Taking into account Eq. (\ref{eq1a}), we find
\begin{equation}
\frac{2}{3}u_{1}v_{1}-\frac{1}{3}\omega u_{2}-\frac{4\pi G}{c^{2}}D=0.
\label{eq10}
\end{equation}

Since $D$ is a physical parameter, it would be natural to use it as {\em
given} and derive one of the other constants. But this complicates
calculations without substantial advantages, as its value depends on the
normalization of the scale factor $a$, which can be fixed arbitrarily.
Thus, we will determine $D$ from Eq. (\ref{eq10})
\begin{equation}
D=\frac{c^{2}}{12\pi G}\left( 2u_{1}v_{1}-\omega u_{2}\right).
\label{eq11}
\end{equation}
Being of course $ D > 0 $, this gives limitations on the choice of $ u_1 $,
$ u_2 $, $ v_1 $.

The well known \cite{Fabris,Bin} solution $\varphi
=2/\sqrt{3(1+\alpha )}\log t$, coming from the potential
\begin{equation}
V(\varphi )=\frac{2(1-\alpha )}{3(1+\alpha )^{2}}\exp \left( -\sqrt{
3(1+\alpha )}\varphi \right) ,
\end{equation}
is a very particular case of what we find in Eqs. (\ref{eq8}) and
(\ref{eq9} ). It can be obtained by setting
\begin{equation}
B^{2}=\frac{2(1-\alpha )}{3(1+\alpha )^{2}}\quad ;\quad \sigma =\sqrt{
3(1+\alpha )}\quad ;\quad u_{2}=v_{2}=v_{1}=0.
\end{equation}

Eq. (\ref{eq11}) then gives $D=0$, so that we get a model without matter,
not really interesting in our context.

A more interesting possibility is given by the choice $u_{1} = v_{1} =
v_{2} = 0$, involving, from Eq. (\ref{eq11}), $u_{2} < 0$. This implies, in
fact,
\begin{equation}
D = - \frac{c^2 \omega u_2}{12 \pi G}, \quad a^3 =
\frac{1}{2}{u_2}^2
\omega t^2, \quad \rho_{\varphi} = \frac{4}{\sigma^2 t^2} \,\,.
\end{equation}
In this particular case, $\rho_{\varphi}$ ($\propto a^{-3}$) scales as
$\rho_m$. (But see below for a discussion on the scaling properties.)

Without any special assumptions on constants, we can get many important
quantities as functions of $u$ and $v$ (we do not write them explicitly in
terms of $t$, for sake of brevity)
\begin{eqnarray}
\rho _{\varphi }(u,v) &=&\frac{1}{\sigma ^{2}}\left( \frac{(\dot{u}v-u\dot{v}
)^{2}}{2u^{2}v^{2}}+\omega \frac{u}{v}\right) ,  \label{eq12a} \\
p_{\varphi }(u,v) &=&\frac{1}{\sigma ^{2}}\left( \frac{(\dot{u}v-u\dot{v}
)^{2}}{2u^{2}v^{2}}-\omega \frac{u}{v}\right) ,  \label{eq12b} \\
w_{\varphi }(u,v) &=&\frac{(\dot{u}v-u\dot{v})^{2}-2\omega
u^{3}v}{(\dot{u} v-u\dot{v})^{2}+2\omega u^{3}v},  \label{eq12c} \\ H(u,v)
&=&\frac{\dot{u}v+u\dot{v}}{3uv},  \label{eq12d} \\
\Omega _{\varphi }(u,v) &=&\frac{(\dot{u}v-u\dot{v})^{2}+2\omega u^{3}v}{(
\dot{u}v+u\dot{v})^{2}},  \label{eq12e} \\
\Omega _{m}(u,v) &=&\frac{24\pi GDuv}{c^{2}(\dot{u}v+u\dot{v})^{2}}=\frac{4u
\dot{u}v\dot{v}-2\omega u^{3}v}{(\dot{u}v+u\dot{v})^{2}},
\end{eqnarray}
the last equality coming from Eq. (\ref{eq11}). It can be easily checked
that $\Omega _{m}+\Omega _{\varphi }=1$.

\subsection{Physical considerations}

We now pass to simplify the situation with a suitable choice of initial
conditions. Of course, the following choice is not the only one possible.
We shall see, anyhow, that it is able to reproduce observational data.
Other more general choices may improve the situation, but we do not treat
them here.

If $t_{in}\equiv -u_{2}/u_{1}$ (which is always possible, being $u_{1}\neq
0$), the scale factor $a$ is zero, and we can show that there is no other
time $t>t_{in}$ when this occurs again. We can thus fix the time origin in
such a way that $a(0)=0$. This condition has to be interpreted just as an
arbitrary choice of the time origin. The real beginning (of physical
meaning) for the model starts a little bit afterwards, at a time $t_{1}$.
This delay is otherwise arbitrary, so that this setting does not seem to
exclude important cases, as said before, and leads to a great
simplification in the formulae. Now, $a(0)=0$ implies $u_{2}=0$ or
$v_{2}=0$, or both. If we set only one of them to zero, we obtain
$\varphi(0)=\infty $ (which could be accepted, but is rather disturbing),
and, most of all, $\Omega_{\varphi }(0)=1$, which would mean an {\em
initial} scalar-field dominated universe, with a neglectable content of
other types of matter. Of course, if we consider the situation in general,
the scalar field does dominate. But, as already mentioned, in our case we
start after decoupling time, when a matter-dominated behaviour seems to be
more natural. So, if we set $u_{2}=v_{2}=0$, we get instead (in a
matter-dominated situation, then)
\begin{equation}
\varphi (0)=-\frac{1}{\sigma }\log \frac{u_{1}}{v_{1}}\quad ;\quad \Omega
_{\varphi }(0)=0\quad ;\quad D=\frac{c^{2}}{6\pi G}u_{1}v_{1}.\quad
\label{eq13}
\end{equation}

We prefer to stick to this choice, so that we have
\begin{equation}
u(t)=u_{1}t\quad ;\quad v(t)=\frac{1}{6}u_{1}\omega t^{3}+v_{1}t.
\label{eq14}
\end{equation}

Let us now define a time scale $t_{s}$ such that $H(t_{s})=1/t_{s}$, which
is of the order of the age of the universe. This leads to consider a
dimensionless time $\tau \equiv t/t_{s}$. From Eqs. (\ref{eq12d}) and (\ref
{eq14}) we get
\begin{equation}
t_{s}^{2}=\frac{6v_{1}}{\omega u_{1}}.  \label{eq15}
\end{equation}

By means of these choices the formulae found above for the relevant
cosmological parameters reduce to
\begin{eqnarray}
\rho _{\varphi } &=&\frac{2(3+4\tau ^{2})}{\sigma ^{2}t_{s}^{2}(1+\tau
^{2})^{2}},  \label{eq16a} \\
p_{\varphi } &=&-\frac{2(3+2\tau ^{2})}{\sigma ^{2}t_{s}^{2}(1+\tau
^{2})^{2}},  \label{eq16b} \\ w_{\varphi } &=&-\frac{3+2\tau
^{2}}{3+4\tau ^{2}},  \label{eq16c} \\ a &=&(u_{1}v_{1}t_{s}^{2}(1+\tau
^{2})\tau ^{2})^{1/3},  \label{eq16d} \\ (1+z)^{3} &=&\frac{\tau
_{0}^{2}(1+\tau _{0}^{2})}{\tau ^{2}(1+\tau ^{2})},
\label{eq16e} \\
H &=&\frac{2(1+2\tau ^{2})}{3t_{s}\tau (1+\tau ^{2})},  \label{eq16f} \\
\Omega _{m} &=&\frac{1+\tau ^{2}}{(1+2\tau ^{2})^{2}}, \label{eq16g}
\end{eqnarray}
where $z\equiv a(\tau _{0})/a(\tau )-1$ is the redshift, and $\tau _{0}$
indicates the present time.

If we define dimensionless pressure and energy density
\begin{equation}
\tilde{p}_{\varphi }\equiv \frac{\sigma ^{2}t_{s}^{2}}{2}p_{\varphi }\quad
;\quad \tilde{\rho}_{\varphi }\equiv \frac{\sigma ^{2}t_{s}^{2}}{2}\rho
_{\varphi },  \label{eq17}
\end{equation}
we find the equation of state for the scalar field
\begin{equation}
\tilde{p}_{\varphi }=\tilde{\rho}_{\varphi }-12+6\sqrt{4-\tilde{\rho}_{\varphi }},
\label{eq18}
\end{equation}
which is well approximated by
\begin{equation}
\tilde{p}_{\varphi }=-0.382\tilde{\rho}_{\varphi }-0.196\tilde{\rho}_{\varphi }^{2}.
\label{eq18a}
\end{equation}

In Fig. 1, we compare the plots of the two functions in Eqs. (\ref{eq18})
and (\ref{eq18a}), and show that the approximation is quite good; a
comparison is also made with a straight line $\tilde{p}_{\varphi }= - 0.86
\tilde{\rho}_{\varphi }$, where the coefficient has been obtained through a
numerical approximation, as well those in Eq. (\ref{eq18a}).

From Eqs. (\ref{eq16a}) and (\ref{eq16d}) it is possible to derive
\begin{equation}
\delta_1 \equiv \frac{d \log \rho_{\varphi}}{d \log a} =
- \frac{3\tau^2(5 + \tau^2)}{(3 + \tau^2)(1 + 2\tau^2)}.
\label{eq19}
\end{equation}
Thus, it is clearly
\begin{equation}
\delta_1 \longrightarrow 0 \quad \mbox{ for } \tau \longrightarrow 0,
\quad \delta_1 \longrightarrow - \frac{3}{2} \quad \mbox{ for } \tau
\longrightarrow \infty,
\label{eq19a}
\end{equation}
so that we asymptotically have {\em two}  scaling regimes:
$\rho_{\varphi}\approx const.$ for early times, and $\rho_{\varphi}
\propto a^{-3/2}$ for late times. In fact, this approximation holds for a
very long time, well far from the asymptotic  values. Indeed, computing
the $n$-th derivative $\delta_n
\equiv a(d\delta_{n-1}/d(\tau^2))/(da/d(\tau^2))$, it is possible to show
that they are all asymptotically zero up to, say, $n = 10$.

An estimate of the moment in which the regime changes can be given finding
the maximum (or the minimum) of $\delta_2$. This is achieved at $\tau\simeq
0.3$, and it is remarkable that this result depends only on $t_s$. All this
discussion shows that it is possible to generalise the concept of scaling
solutions. The situation is illustrated in Figs. (2) and (3).

As a matter of fact, in the literature it is widely accepted that using an
exponential potential leads to a dark energy density which scales like
matter. Our results seem to be in contrast with this statement, which is a
consequence of assuming $w_{\varphi}$ almost perfectly constant. It is not
our case, as shown in the following.

As in \cite{SWZ}, we can use the function $\Gamma \equiv V''V/(V')^2$.
Defining
\begin{equation}
x \equiv \frac{{\dot{\varphi}}^2}{2V} = \frac{1 + w_{\varphi}}{1 -
w_{\varphi}}, \quad  \dot{x} \equiv \frac{d \log x}{d \log a},
\quad \ddot{x} \equiv \frac{d^2 \log x}{d \log a^2}\,,
\label{eq19b}
\end{equation}
it is possible to find \cite{SWZ}
\begin{equation}
\Gamma = 1 + \frac{w_m - w_{\varphi}}{2(1 + w_{\varphi})}
- \frac{1 + w_m - 2w_{\varphi}}{2(1 + w_{\varphi})}\frac{\dot{x}}{6 + \dot{x}}
- \frac{2}{1 + w_{\varphi}}\frac{\ddot{x}}{(6 + \dot{x})^2} \,.
\label{eq19c}
\end{equation}
If one makes the assumption that $w_{\varphi}$ is nearly constant, then
$\dot{x} \simeq \ddot{x} \simeq 0$. Since it is $\Gamma = 1$ strictly in
the case of our potential, Eq. (\ref{eq19c}) implies $w_m \approx
w_{\varphi}$. The point is that it is not true that our exact solutions
for the exponential potential lead to $\dot{x} \simeq \ddot{x} \simeq 0$.
For instance, we have
\begin{equation}
x = \frac{\tau^2}{3(1 + \tau^2)} , \quad \dot{x} = \frac{3}{1 +
2\tau^2}\,,
\end{equation}
so that $\dot{x} \longrightarrow 0$ only asymptotically. Since in our
model $\tau \lesssim 1$, $x$ and $\dot{x}$ are then far from being zero.

But let us also consider the third term in the right-hand side of Eq.
(\ref{eq19c}) with $w_m = 0$ (matter is simply dust in our model).
Substituting our solution, we get
\begin{equation}
- \frac{1 + w_m - 2w_{\varphi}}{2(1 + w_{\varphi})}\frac{\dot{x}}{6 + \dot{x}}
= \frac{9 + 8\tau^2}{12\tau^2 + 16\tau^4}\,;
\end{equation}
we see that this expression diverges for $\tau \longrightarrow 0$, being
always $> 0.5$ in the useful range of $\tau$.

In our opinion, the main check for the solution in Eq. (\ref{eq14}) is
thus only its capability to reproduce the experimental results, which we
are going to do just below.

From Eq. (\ref{eq16g}) we get
\begin{equation}
{\tau
_{0}}^2=\frac{1-4{\Omega_m}_{0}+\sqrt{1+8{\Omega_m}_{0}}}{8{\Omega_m}_{0}}.
\label{eq20}
\end{equation}

Once we give an acceptable value for ${\Omega _{m}}_{0}$, we obtain a
value for $\tau _{0}$. For instance, ${\Omega _{m}}_{0}=0.3$ gives $\tau
_{0}=0.82$, and this implies ${w_{\varphi }}_{0}=-0.76$. If the value
$H_{0\text{ }}=100h\ km\ s^{-1}Mpc^{-1}$ is also given, we get
\begin{equation}
t_{s}=\frac{2(1+2\tau _{0}^{2})}{3H_{0}\tau _{0}(1+\tau _{0}^{2})}=\frac{
1.\,\allowbreak 14}{H_{0}};  \label{eq21}
\end{equation}
assuming $h=0.7$, we have $t_{s}=15.8{\times} 10^{9}$ years, and $t_{0}=13{{\times}
}10^{9}$ years. It is also possible to obtain the relation between
${w_{\varphi }}_{0}$ and ${\Omega _{m}}_{0}$
\begin{equation}
{w_{\varphi }}_{0}=\frac{1+8{\Omega _{m}}_{0}-3\sqrt{1+8{\Omega
_{m}}_{0}}}{ 4(1-{\Omega _{m}}_{0})}.  \label{eq22}
\end{equation}
For ${\Omega _{m}}_{0}=0.2\div 0.4$, we get ${w_{\varphi
}}_{0}=-0.\,\allowbreak 699\,\div -0.\,\allowbreak 811$, and the value
$-0.5$ is reached only in the case of ${\Omega
_{m}}_{0}=0$. It is also possible to obtain $w_{\varphi }$ as a function
of the redshift
\begin{equation}
w_{\varphi }=-\frac{2\zeta +\sqrt{\zeta (4+\zeta )}}{\zeta +2\sqrt{\zeta
(4+\zeta )}},  \label{eq23}
\end{equation}
where $\zeta \equiv (1+z)^{3}$.

Figs. 4, 5, and 6 show the dependance of $ w-\phi $ versus $ \tau $,
$\Omega_{m0}$, and $ \zeta $, respectively. We can see that $w_{\varphi }$
varies much and has values $\sim
-0.78$. For $\zeta
=1$ (now) it is $w_{\varphi }=-0.77$, and already for $\zeta
=4$ ($z\simeq 0.59$) we find $w_{\varphi }\simeq -0.89$. It is remarkable
that the values of $|w_{\varphi}|$ are greater than $0.7$. According to our
knowledge, this feature is found only in \cite{Brax}.

Another interesting quantity is the present value of $\varphi $. After
staightforward algebra we get
\begin{equation}
\varphi _{0}=-\frac{1}{\sigma }\log \left( \frac{27H_{0}^{2}}{16B^{2}}(1-4{
\Omega _{m}}_{0}+\sqrt{1+8{\Omega _{m}}_{0}})\right) ,  \label{eq24}
\end{equation}
and we see that this value depends on the observed parameters and on the
value of $B^{2}$, which was until now completely undetermined. Now, for
$
\tau _{0}=0.82$ and $t_{0}=13{{\times} }10^{9}$ years, we have $B^{2}=2.5{
{\times} }10^{-47}\exp (-\sigma \varphi _{0}) \,GeV^{4}$. Considering $\varphi
_{0}\approx 1/6\ M_{P}$ ($M_{P}$ being the Planck mass) we see that $\exp
(-\sigma \varphi _{0})\approx 1$, and we can determine the unknown
parameter for the potential
\begin{equation}
V(0)\equiv B^{2}\approx 2.5{{\times} }10^{-47} \,GeV^{4}.
\end{equation}

But we have also to observe that a \ `little' change in $\varphi _{0}$
entails a `large' change in $B^{2}$. For instance, if $\varphi _{0}\approx
M_{P}$, then $\exp (-\sigma \varphi _{0})\approx 0.0025$ and $B^{2}$
changes of three orders of magnitude. Due to Eq. (\ref{eq13}), we have
that $\varphi (0)=-1/\sigma \log (2\pi GB^{2}t_{s}^{2}/c^{2})$. This means
therefore that a relatively wide range of initial values of $\varphi $
ends up to a narrower set of final $\varphi _{0}$'s.

Thus, everything seems to work fine, but things are more complicated.
Indeed, one has to ask what is really measured in the supernovae experiment.
The value ${\Omega _{m}}_{0}=0.3$ is not a direct consequence of the data,
since it depends on the model, which uses the constant $\Lambda $-term. What
we really measure is the distance modulus, so that it is this quantity that
we should compare in the two situations. Here, we limit ourselves to a very
qualitative discussion.

Let us recall, then, the definitions of luminosity distance (in Mpc)
\begin{equation}
d_{L}=3000(1+z)\int_{0}^{z}\frac{dz
{\acute{}}%
}{H(z%
{\acute{}}%
)},  \label{eq25}
\end{equation}
and distance modulus
\begin{equation}
\delta \equiv m-M=5\log _{10}d_{L}(z)+25.  \label{eq26}
\end{equation}

We have thus to compare this last quantity in the case when $H(z)$  is
taken from the usual model with $\Lambda $ \cite{Perl3,Jha}, that is,
\begin{equation}
H(z)=H_{0}\sqrt{(1+z)^{2}(1+{\Omega_m}_{0}z)-z(2+z)(1-{\Omega_m}_{0})},
\label{eq27}
\end{equation}
with the one obtained eliminating $\tau $ from Eqs. (\ref{eq16e}) and
(\ref {eq16f}).

In Fig. 7 we compare $\delta $ with $\tilde{\delta}$ (let us mark with a
$\sim$ the values for the model with $\Lambda $). The agreement is almost
perfect, up to $0.06\%$. But there is a trick! $\tilde{
\delta}$ was obtained from a value ${\tilde \Omega}_{m0}=0.37$ . Of course,
this value is still in the possible range but at its limit. If we decide
to trust strongly on the value ${\tilde \Omega}_{m0}=0.3$ and want to
obtain the same good agreement, we have to change the value of $\tau
_{0}$ to $1.22$. This gives a very different value ${\Omega
_{m}}_{0}=0.16$ in the model with $\varphi $. This is again at the limit
of possible estimates (due to other investigations on dark matter).

In conclusion, we see that this solution (with the potential in Eq. (\ref
{eq4})) is indeed difficult to fully adapt to observed data, but for reasons
which are not easy to investigate without general exact solutions. Moreover,
it is not clearly incompatible (until we get better data); therefore, it
seemed to us useful to present it in detail.

\section{Two exponentials combined}

\subsection{Mathematical treatment}

We now consider a combination of two exponentials, which will give us much
better results, as expected. The procedure strictly follows the above one.

Let us consider the potential
\begin{equation}
V(\varphi )=A^{2}e^{\sigma \varphi }+B^{2}e^{-\sigma \varphi },
\label{eq2.1}
\end{equation}
with $\sigma ^{2}=12\pi G/c^{2}$ as before, and $A^{2}$, $B^{2}$ arbitrary
parameters. We use, now, the following change of variables
\begin{equation}
a^{3}=\frac{u^{2}-v^{2}}{4}\quad ;\quad \varphi =\frac{1}{\sigma }\log
\frac{ B(u+v)}{A(u-v)},  \label{eq2.2}
\end{equation}
which is invertible, provided that $a\neq 0$. This leads to
\begin{equation}
V(u,v)=2AB\frac{u^{2}+v^{2}}{u^{2}-v^{2}}.  \label{eq2.3}
\end{equation}

With these variables Eqs. (\ref{eq2a}) and (\ref{eq3a}) are rewritten as
\begin{equation}
\ddot{u}=\omega ^{2}u\quad ;\quad \ddot{v}=-\omega ^{2}v,  \label{eq2.4}
\end{equation}
where now
\begin{equation}
\omega ^{2}=\frac{12\pi GAB}{c^{2}}.  \label{eq2.5}
\end{equation}

Again, the integration is immediate, and gives the general solutions
\begin{eqnarray}
u(t) &=&\alpha e^{\omega t}+\beta e^{-\omega t},  \label{eq2.6} \\
v(t) &=&v_{1}\sin (\omega t+v_{2}),  \label{eq2.7}
\end{eqnarray}
with $\alpha $, $\beta $, $v_{1}$, $v_{2}$ arbitrary constants. As before,
we derive $D$ from the constraint in Eq. (\ref{eq1a})
\begin{equation}
D=-\frac{c^{2}\omega ^{2}(v_{1}+4\alpha \beta )}{24\pi G}.  \label{eq2.8}
\end{equation}

Being $D>0$, this implies $v_{1}<-4\alpha \beta $. A change in the sign of
$ v_{1}$ has the only effect of changing the sign of $\varphi $, and
interchanging $A^{2}$ with $B^{2}$. So, we can set $v_{1}>0$ without any
loss of generality (the case $v_{1}=0$ is obviously equivalent to
considering a $\Lambda $-term). As a consequence, $\alpha $ and $\beta $
must be non zero and with opposite signs.

Again, we can write some important functions in terms of $u$, $v$
\begin{eqnarray}
\rho _{\varphi } &=&\frac{2((\dot{v}u-\dot{u}v)^{2}+\omega ^{2}(u^{4}-v^{4}))
}{\sigma ^{2}(u^{2}-v^{2})^{2}},  \label{eq2.19} \\ p_{\varphi }
&=&\frac{2((\dot{v}u-\dot{u}v)^{2}-\omega ^{2}(u^{4}-v^{4}))}{
\sigma ^{2}(u^{2}-v^{2})^{2}},  \label{eq2.20} \\
w_{\varphi } &=&\frac{(\dot{v}u-\dot{u}v)^{2}-\omega ^{2}(u^{4}-v^{4})}{(
\dot{v}u-\dot{u}v)^{2}+\omega ^{2}(u^{4}-v^{4})}.
\end{eqnarray}

(This last expression gives $w_{\varphi }\simeq -1$ when
$(\dot{v}u-\dot{u}v)^{2}\ll u^{4}-v^{4}$, which certainly can happen for
sufficiently large times.)

For the redshift we have
\begin{equation}
(1+z)^{3}=\frac{4{a_{0}}^3}{u^{2}-v^{2}},  \label{eq2.22}
\end{equation}
and finally
\begin{eqnarray}
H(u,v) &=&\frac{2(u\dot{u}-v\dot{v})}{3(u^{2}-v^{2})},  \label{eq2.23} \\
\Omega _{m}(u,v) &=&-\frac{(u^{2}-v^{2})(\dot{v}^{2}-\dot{u}^{2}+\omega
^{2}(u^{2}+v^{2}))}{(u\dot{u}-v\dot{v})^{2}},  \label{eq2.24} \\
\Omega _{\varphi }(u,v) &=&\frac{(\dot{v}u-\dot{u}v)^2 +\omega
^{2}(u^{4}-v^{4})}{(u\dot{u}-v\dot{v})^{2}},  \label{eq2.25}
\end{eqnarray}
with $\Omega _{m}+\Omega _{\varphi }=1$, of course.

\subsection{Physical considerations}

The potential in Eq. (\ref{eq2.1}) has a nonzero minimum $V_{min} = A^2 +
B^2$. This is unusual in quintessence theory, since $V_{min} = 0$ is
optimal to remove the fine-tuning problem. Indeed, $V_{min} \neq 0$ can be
seen as a disguised $\Lambda$-term. Anyway, in the following we find that
this is actually the case when the scalar field is almost stationary near
the minimum of the potential. But there is also the possibility of a slower
rolling far from the minimum. The situation is then similar to that in Sec.
II, but the additional term in the potential now allows to achieve a better
agreement with observational data.

Let us now make a trial for the choice of the free parameters. We set
again $a(0)=0$, and ask for nonsingular $\varphi (0)$. The situation, and
hence the interpretation, is the same as above. Thus, we pose $\alpha
=-\beta
=\lambda
/2 $, $v_{2}=0$. It is also possible to fix an arbitrary normalization for $
a $ and set $v_{1}=1$, obtaining at last
\begin{equation}
u(t)=\lambda \sinh (\omega t)\quad ;\quad v(t)=\sin (\omega t).
\label{eq2.26}
\end{equation}

We get now
\begin{equation}
D=\frac{c^{2}\omega ^{2}(\lambda ^{2}-1)}{24\pi G},  \label{eq2.27}
\end{equation}
implying $|\lambda |>1$, and
\begin{equation}
\varphi (0)=\frac{1}{\sigma }\log \frac{B(\lambda +1)}{A(\lambda -1)}.
\label{eq2.28}
\end{equation}

We define a dimensionless time $\tau =\omega t$ and get
\begin{eqnarray}
a(\tau ) &=&\left( \frac{\lambda ^{2}\sinh ^{2}\tau -\sin ^{2}\tau }{4}
\right) ^{1/3},  \label{eq2.29} \\
(1+z)^{3} &=&\frac{\lambda ^{2}\sinh ^{2}\tau _{0}-\sin ^{2}\tau _{0}}{
\lambda ^{2}\sinh ^{2}\tau -\sin ^{2}\tau },  \label{eq2.30} \\
H(\tau ) &=&\frac{\omega (\sin (2\tau )-\lambda ^{2}\sinh (2\tau ))}{3(\sin
^{2}\tau -\lambda ^{2}\sinh ^{2}\tau )},  \label{eq2.31} \\
w_{\varphi }(\tau ) &=&\frac{\lambda ^{2}(\cosh \tau \sin \tau -\cos \tau
\sinh \tau )^{2}+(\sin ^{4}\tau -\lambda ^{4}\sinh ^{4}\tau )}{\lambda
^{2}(\cosh \tau \sin \tau -\cos \tau \sinh \tau )^{2}-(\sin ^{4}\tau
-\lambda ^{4}\sinh ^{4}\tau )}, \\
\Omega _{m}(\tau ) &=&\frac{2(\lambda ^{2}-1)(\cos (2\tau )+\lambda
^{2}\cosh (2\tau )-1-\lambda ^{2})}{(\sin (2\tau )-\lambda ^{2}\sinh (2\tau
))^{2}}.
\end{eqnarray}

In comparison with the situation in Sec. II, we now have one more free
parameter. This gives the possibility of a much better agreement with
observational data.

We can also take into some account the final and initial values of
$\varphi $, i.e., $\varphi _{0}$ and $\varphi _{i}=\varphi (0)$. If
$\tau_{0}$ is the present dimensionless time, we get
\begin{equation}
\exp \left( \sigma (\varphi _{i}-\varphi _{0})\right) =\frac{\lambda
^{2}\sinh \tau _{0}+\lambda (\sin \tau _{0}-\sinh \tau _{0})-\sin \tau _{0}}{
\lambda ^{2}\sinh \tau _{0}-\lambda (\sin \tau _{0}-\sinh \tau _{0})-\sin
\tau _{0}}.
\end{equation}

When $\lambda >>1$, we have $\varphi _{i}\simeq \varphi _{0}$ . In fact, $
\varphi $ is practically constant and we have $w_\varphi \simeq -1$, with nearly
perfect emulation of a cosmological constant. On the other hand, if $\lambda
\simeq 1$, then $\exp \left( \sigma (\varphi _{i}-\varphi _{0}\right)
)\simeq 0$. This can be interpreted as $\varphi _{i}\simeq \infty $, or,
better, as the possibility of a wide range of $\varphi _{i}$'s, with
nearly the same final $\varphi _{0}$.

Whatever is $\lambda $, anyhow, it is possible to obtain a good agreement
with observational data. We here give only two extreme cases, with
$\lambda =30$ and $\lambda =1.1$.

In the first case (case I) we set
\begin{equation}
\lambda =30\quad ;\quad \tau _{0}=1.2\quad ;\quad \omega = 2.8 {{\times}}10^{-18} s^{-1} =
2.8 {{\times}} 10^{-42} \,GeV,
\end{equation}
which gives
\begin{equation}
H_{0}=70\quad ;\quad {\Omega _{m}}_{0}=0.3\quad ;\quad {w_{\varphi }}
_{0}=-0.999.
\end{equation}

In the second case (case II) posing
\begin{equation}
\lambda =1.1\quad ;\quad \tau _{0}=0.44\quad ;\quad \omega =1.07{{\times}}10^{-18} s^{-1} =
1.07 {{\times}} 10^{-42} \,GeV
\end{equation}
gives
\begin{equation}
H_{0}=70\quad ;\quad {\Omega _{m}}_{0}=0.3\quad ;\quad {w_{\varphi }}
_{0}=-0.76.
\end{equation}

Again, we find that the value of $w_{\varphi}$ is less than $- 0.7$, as in
Sec. II. In Figs. 8 and 9 the distance modulus $\delta $ for these two
cases is compared with the $\Lambda $-term case. As before, the agreement
is quite good, but now we have ${\tilde\Omega}_{m0}= 0.30$, so that, with
this very rough analysis, it is impossible to make a distinction.

Let us present again the plots of $\log (\rho )$ versus $\log (a)$. They
are shown in Figs. 10 and 11. We see that the case I is practically
indistinguishable from a $\Lambda $-term, while the case II is more
similar to the situation in Sec. II, with different scaling regimes. It is
interesting to note that the regimes are now three, even if the last one
seems to be important only in the remote future. We stress that also in
this case, as well as in Sec. II, $w_{\varphi}$ is not constant.

As a final result, we plot in Figs. 12 and 13 the equation of state for
the scalar field. Now it is impossible to show an exact analytical
expression, so that we only give the plots (in arbitrary units) in the two
examined cases. It is interesting to note that in the case I, although so
similar to the pure cosmological constant case, we nonetheless obtain a
nontrivial plot for the equation of state. But, clearly, this point
deserves further investigation.

\section{Conclusions}

We have discussed two particular kinds of potentials which have allowed
the general exact integration of Friedmann equations in presence of dust
(ordinary and cold dark matter) and scalar field. This has been achieved
by performing suitable transformations of variables. Such transformations
have not been guessed by chance, but are the results of a well known
procedure, the {\em N\"{o}ther Symmetry approach} \cite{deR2,RivNC,deR3,deR1},
based on an action principle. This was not mentioned before because it was
unnecessary to the main goal of the discussion. Nevertheless, it now seems
appropriate to stress the power of such a procedure, which allows to solve
cosmological equations, often giving also informations on the potential
and/or the possible coupling between the scalar field and the curvature of
spacetime, without any limitation on the validity of the solution itself.
(For details, see the literature quoted above, and the references
therein.)

We have seen that, with a suitable choice of integration constants for
both potentials, it is possible to reproduce the main recent results from
supernovae (initially interpreted in a $\Lambda $-model), with
considerable precision, especially in the case II of the second potential.
(The case I does not add much to what is already known in a constant
$\Lambda$-term model.) What is interesting, in our opinion, is that such
kinds of models can bring to a different evaluation of an important
quantity like ${\Omega }{_{m}}_{0}$. This, in a certain sense, sheds new
light on the exponential potential of the first type, which is usually not
considered as completely adequate for quintessence, for instance. But we
have seen that, without considering {\em a priori} $w_{\varphi }$ as a
constant and having a general exact solution, something else can be
learned. As a matter of fact, not having an almost constant $w_{\varphi}$
makes it impossible to treat the tracker condition in the usual way. Also,
we get appreciable values of $w_{\varphi}$, i.e., surely less than $-
0.7$.

Of course, all our discussion is in part still {\em qualitative}, in that
we should need to make a more punctual analysis of observational data, and
verify the best fit with the various models, in order to see whether and
when real differences arise. Anyway, our analysis already seems to confirm
some of the considerations made in \cite{Maor}.

Another important point to note is that, to be realistic and cover the
whole (or, at least, a substantially wider) range of the life of the
universe, radiation (and hot dark matter) must be added into the game.
This could allow to study the CMBR spectrum and the formation of
structures. But it presumably destroys the possibility to integrate the
system of the cosmological equations, leading to the necessity of using
the results we established here only as a guide for a more complete
analysis.

\section*{Acknowledgments}

We thank  Prof. M. Demianski for his useful suggestions and Prof. M. Sazhin
for revision of the manuscript. We also thank Prof. J. D. Barrow for his
comments, and Prof. V. Sahni, who has kindly attracted our attention on
some papers. 

This work has been in part financially sustained by the M.U.R.S.T.
grant PRIN2000 ``SIN.TE.SI.''.

\vfill\eject

{\bf \large Figure Captions}
\bigskip

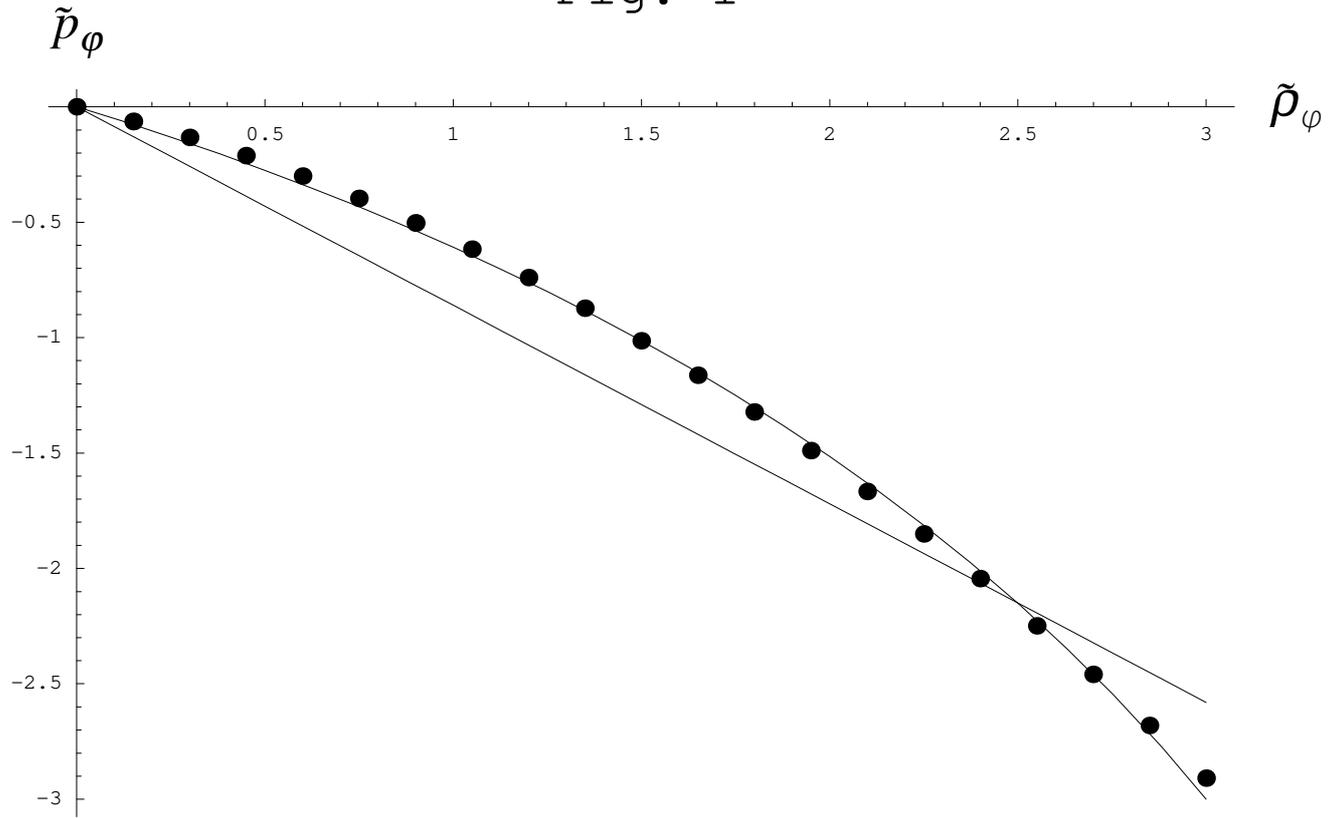
Fig. 1 - Comparison of the state equation in Eq. (36) with an  approximate
quadratic equation  and with an approximate straight line.

\medskip
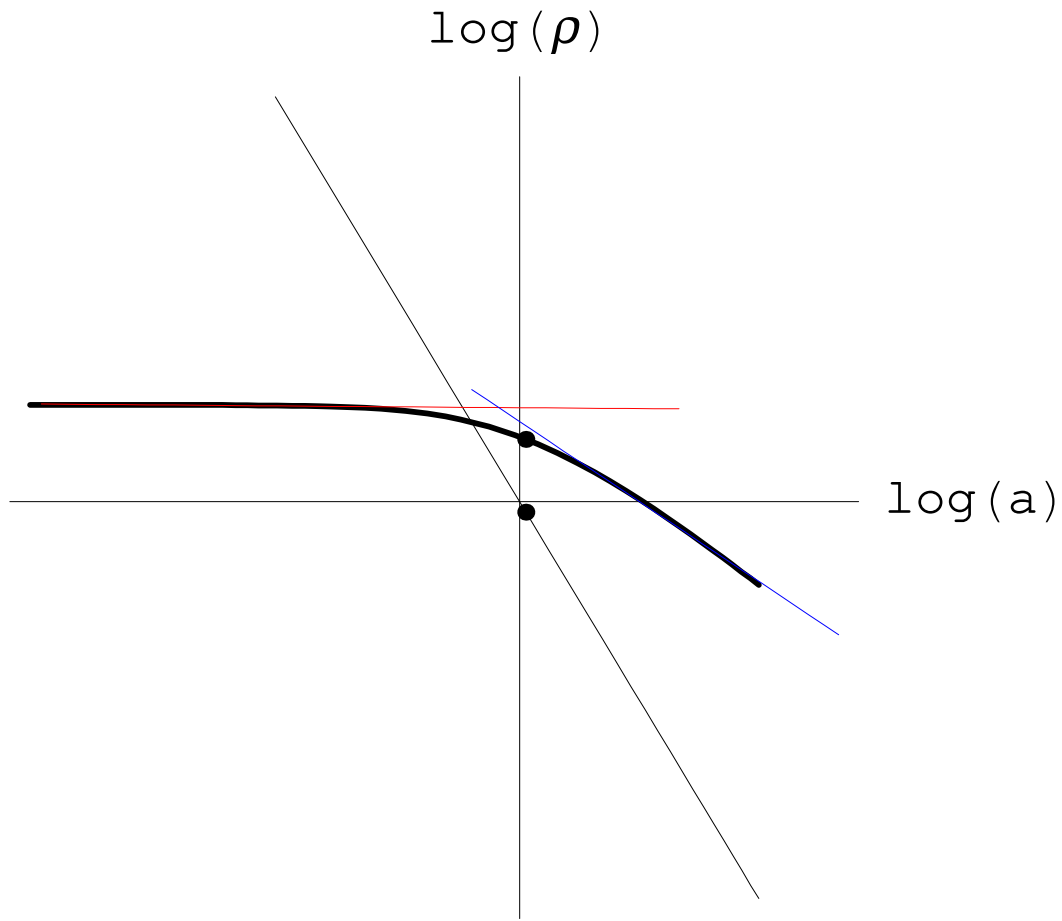
Fig. 2 - The straight line represents the dependence of $ \rho_m $ versus $
a $ ($ \rho_m = D a^{-3}$). The thick curve is for the scalar field. The
two tangents show the asymptotic scaling behaviour in the period of
dominating matter and scalar field, respectively. The bullets indicate
present time, according to $ a = 0.82 $. Units are arbitrary.

\medskip
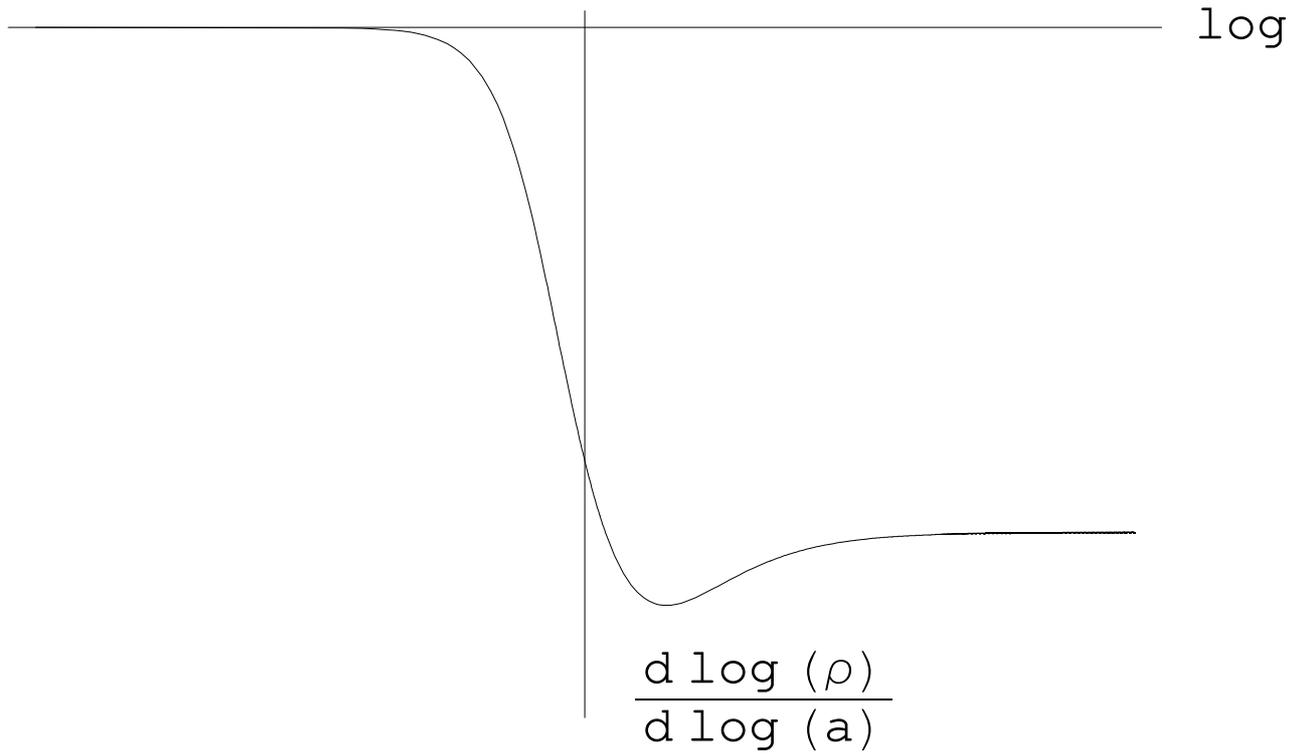
Fig. 3 - The derivative of  $ \log (\rho) $ with respect to $ \log (a) $,
shows a quick transition from one scaling regime to another one. Units are
arbitrary.

\medskip
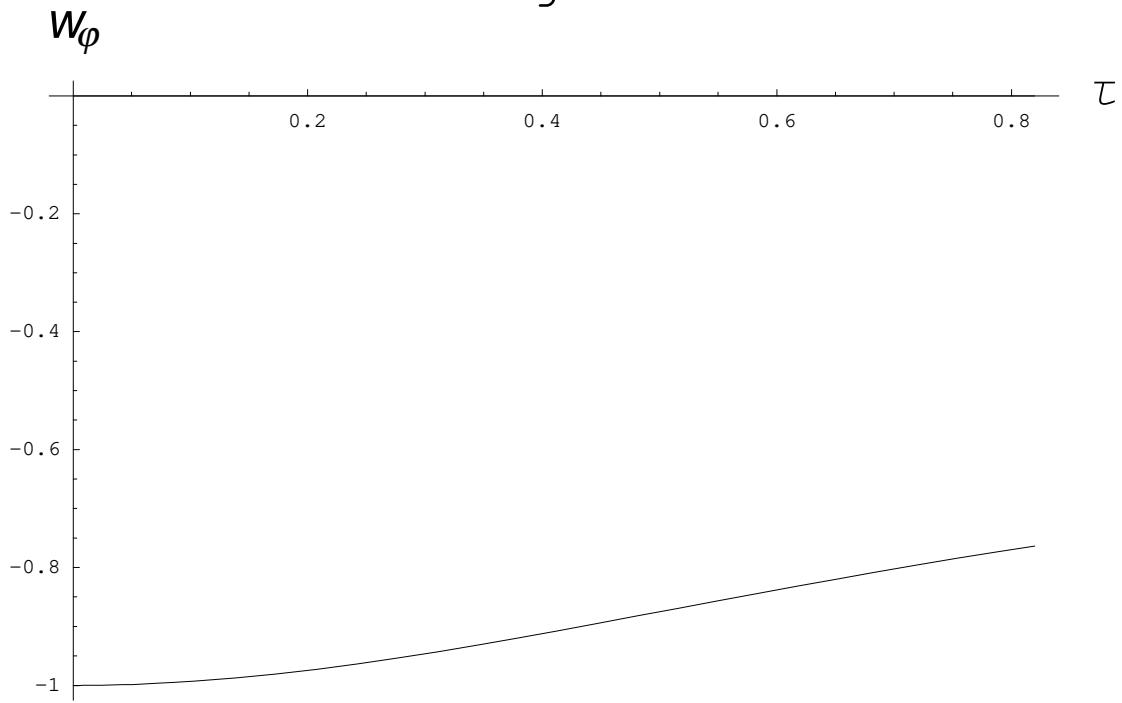
Fig. 4 - The plot of $w_\phi $ versus time shows that $w_\phi $  is far
from being constant.

\medskip
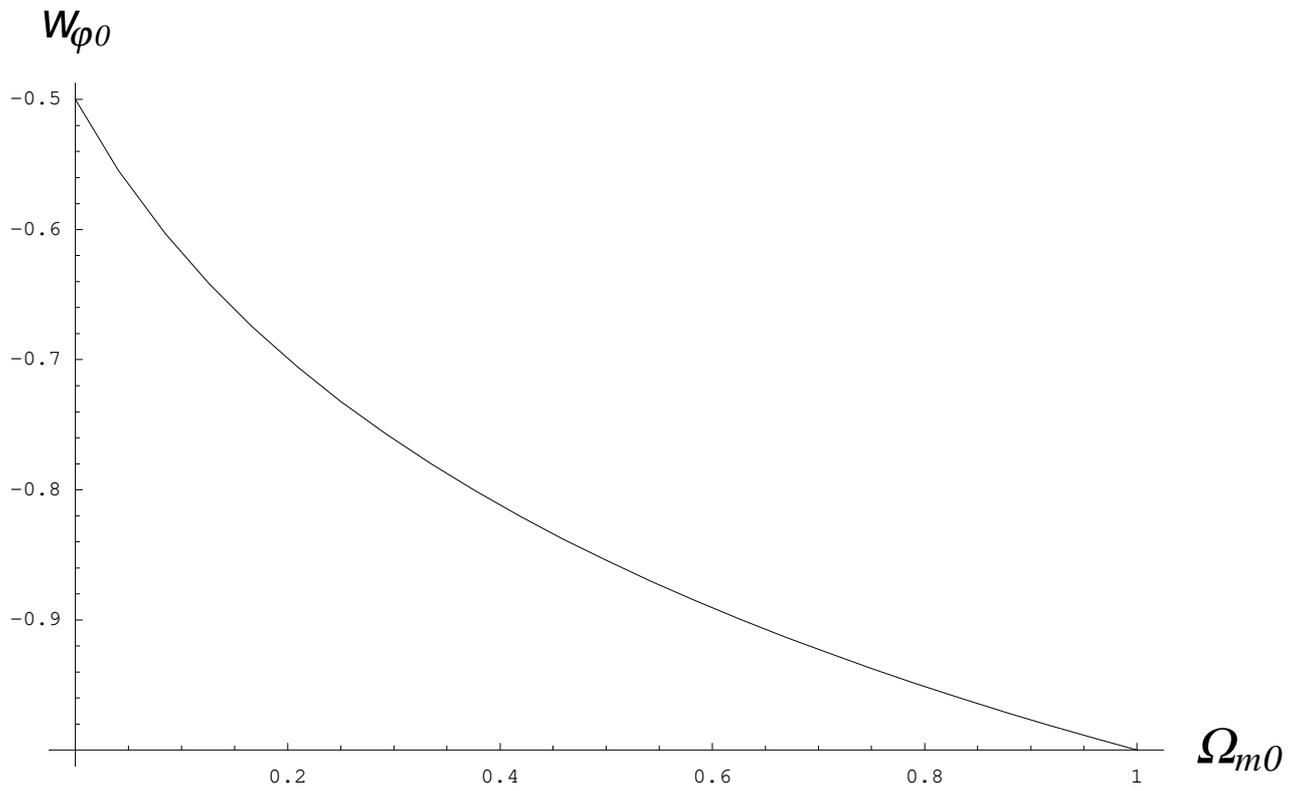
Fig. 5 - $w_{\phi 0}$  versus $ \Omega_{m0} $.

\medskip
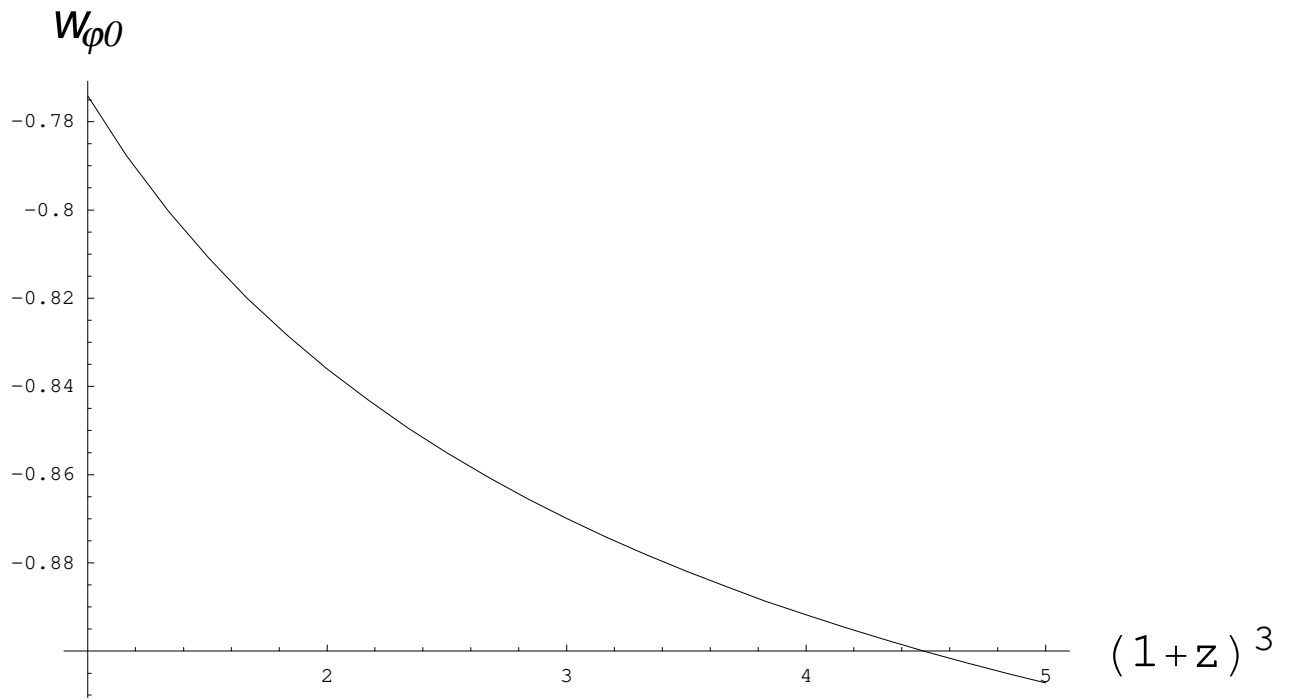
Fig. 6 - This plot shows the dependence of $w_\phi $ from the redshift.

\medskip
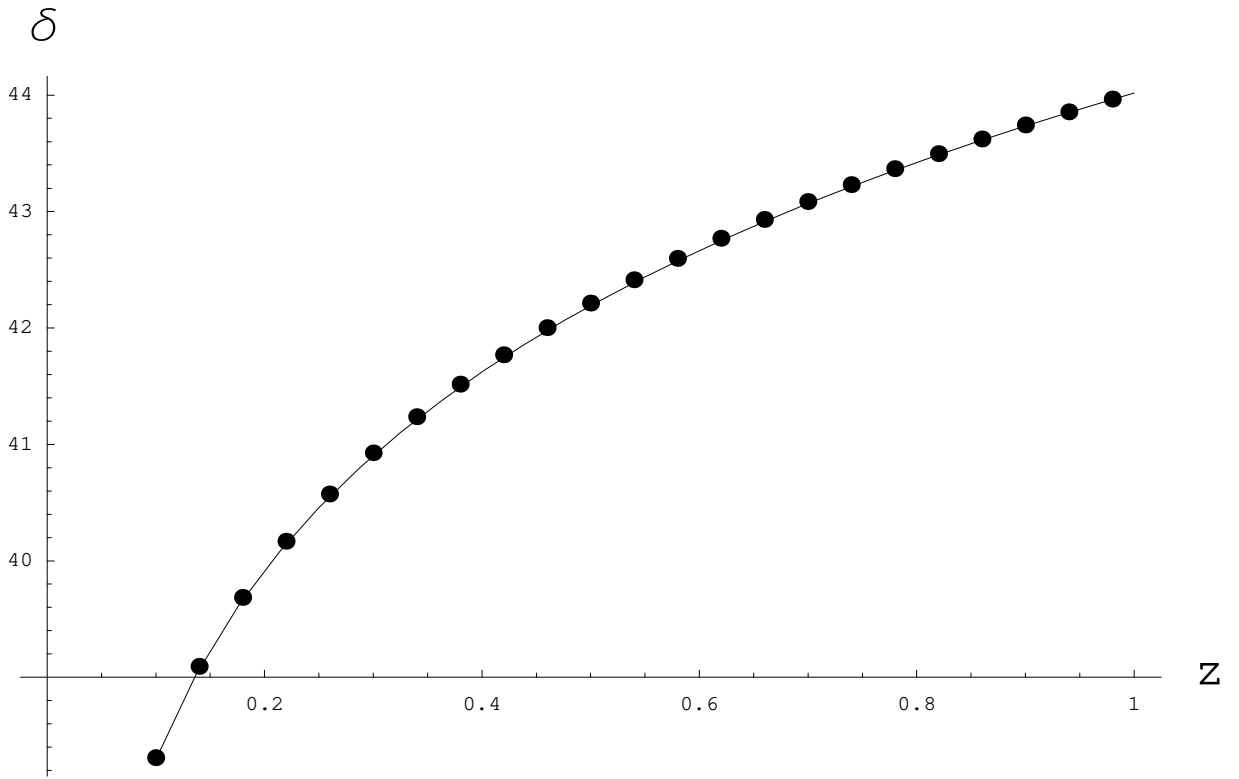
Fig. 7 - Comparison of the distance modulus $ \delta $(derived from Eqs.
(32), (33) and (34),  and assuming $ \Omega_{m0} = 0.3$) (continuous line)
with $\tilde \delta $ (derived from a $\Lambda$-term model,assuming a $
{\tilde \Omega}_{m0} = 0.37)$ (dots).

\medskip
Fig.8 - Comparison of the distance modulus $\delta]$ (derived from Eqs.
(68), (69) and (71) in case I,  and assuming $ \Omega_{m0} = 0.3$)
(continuous line) with $\tilde \delta $  (derived from a $\Lambda$-term
model, assuming $ {\tilde \Omega}_{m0} = 0.3)$) [dots].

\medskip
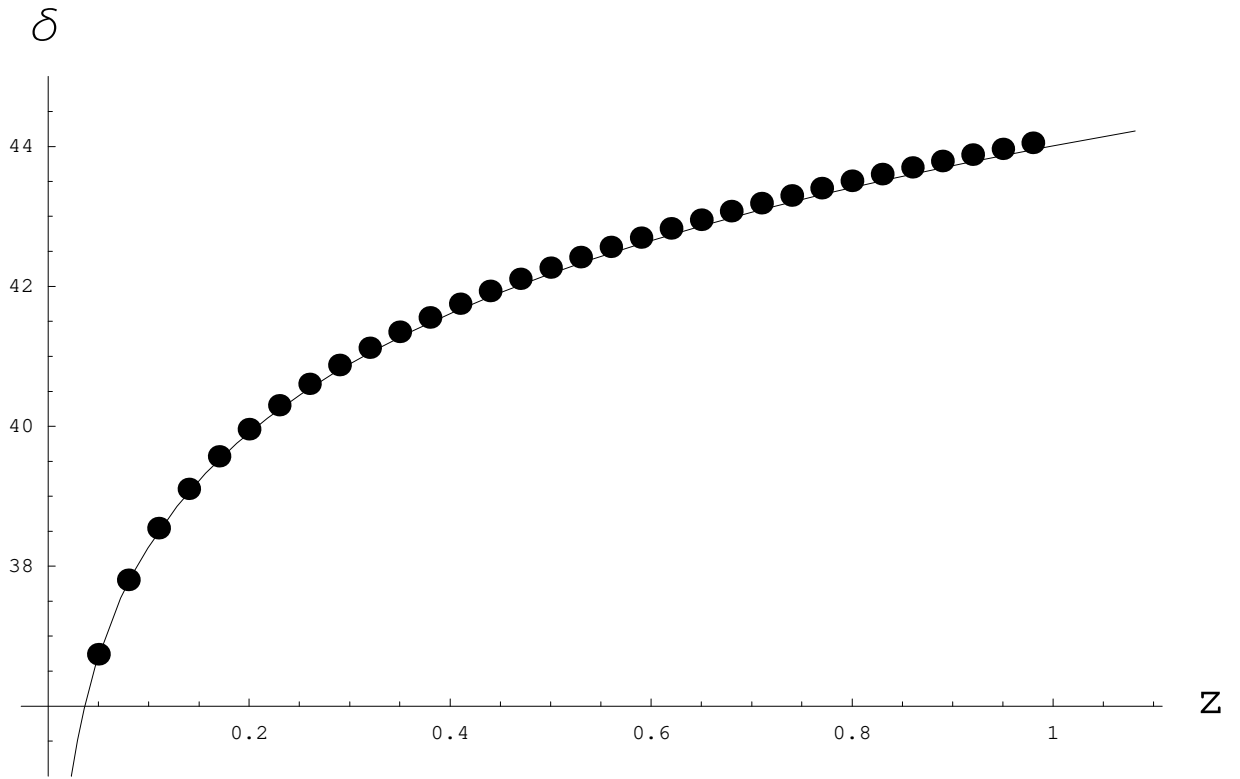
Fig.9 - Comparison of the distance modulus $\delta $(derived from Eqs.
(68), (69) and  (71) in case II,  and assuming $ \Omega_{m0} = 0.3$)
(continuous line) with a  (derived from a $\Lambda$-term model, assuming $
{\tilde \Omega}_{m0} = 0.3)$) [dots].

\medskip
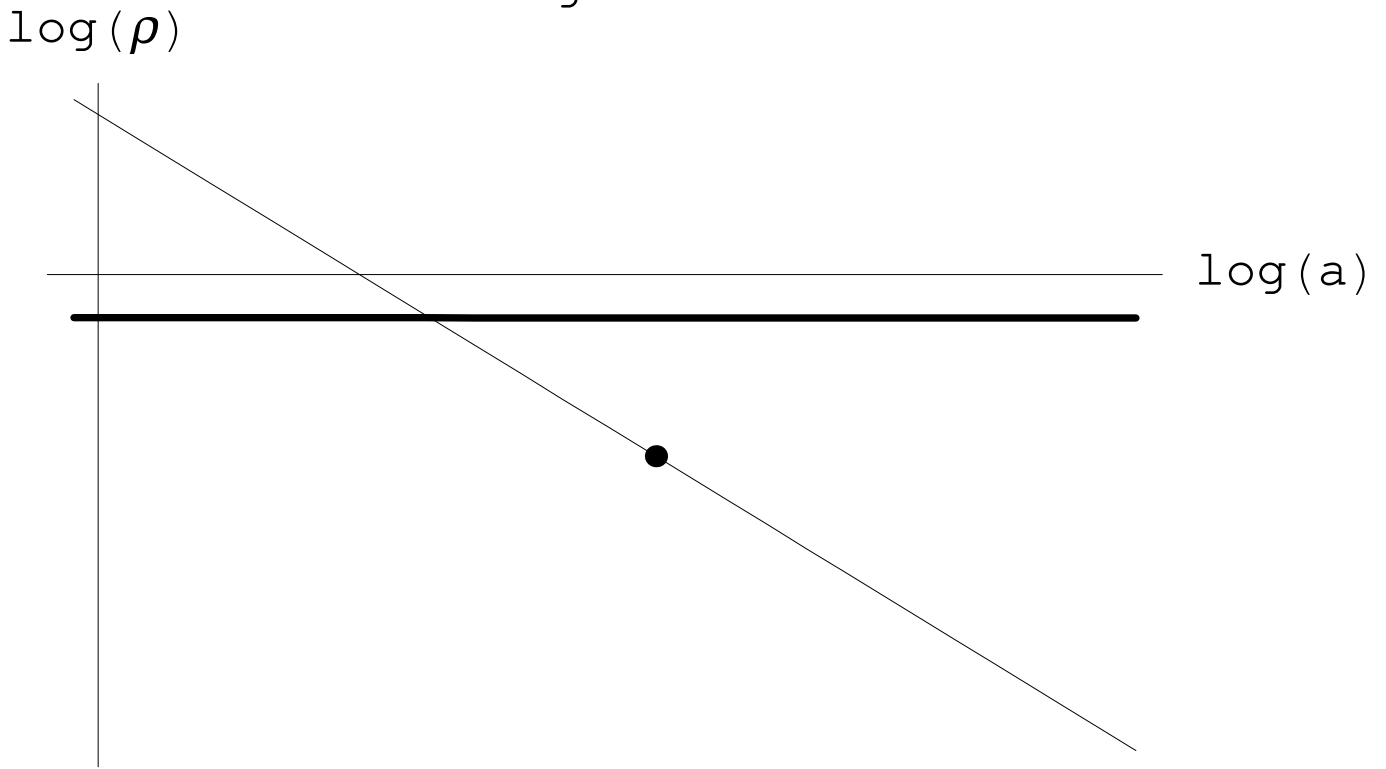
Fig. 10 - Case I. The  straight line  with slope represents the dependance
of $ \log \rho_m $ versus $ \log a $ ($\rho_m  =  D a^{-3}$). The
horizontal thick line is for the scalar field.  It emulates a constant. The
bullet indicates present time, according to $\tau_0 = 1.2$. Units are
arbitrary.

\medskip
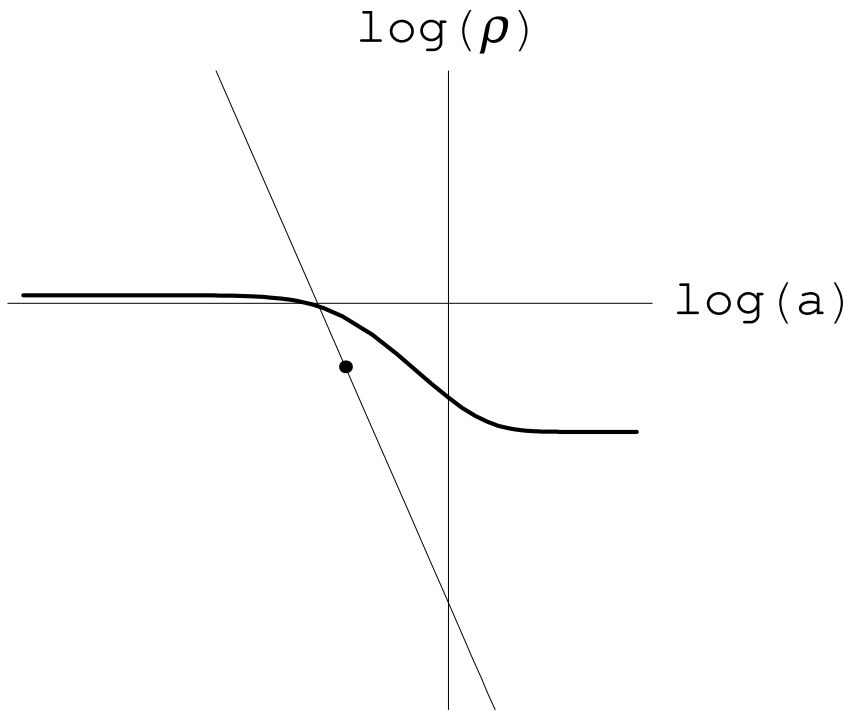
Fig. 11 - Case II. The  straight line  with slope represents the dependance
of $ \log \rho_m $  versus $ \log a $ ($\rho_m  =  D a^{-3}$). The thick
line is for the scalar field.  There are three approximate scaling regimes.
The bullet indicates present time, according to $\tau_0 = = 0.44 $. Units
are arbitrary.

\medskip
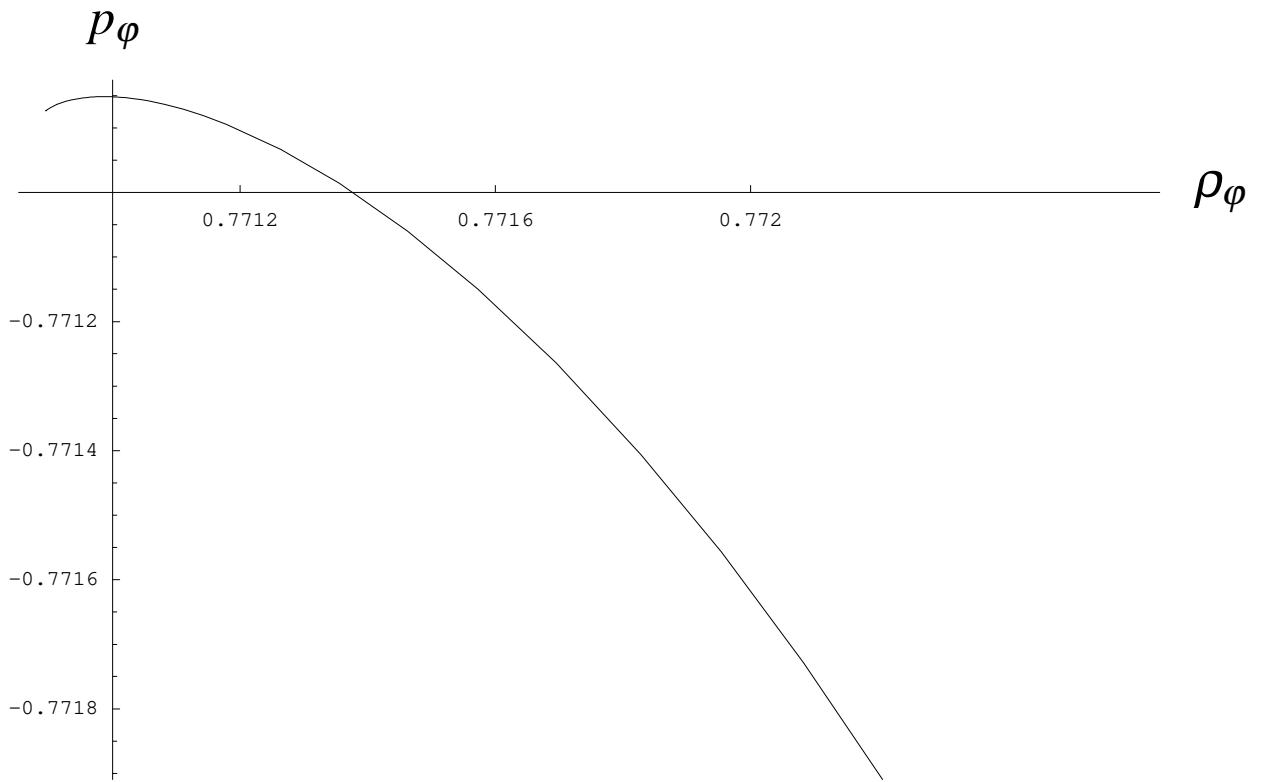
Fig.12 - Case I. State equation for the scalar field. Although it perfectly
emulates a cosmological constant, the equation  is nontrivial. Units are
arbitrary, but coherent with Fig. 13.

\medskip
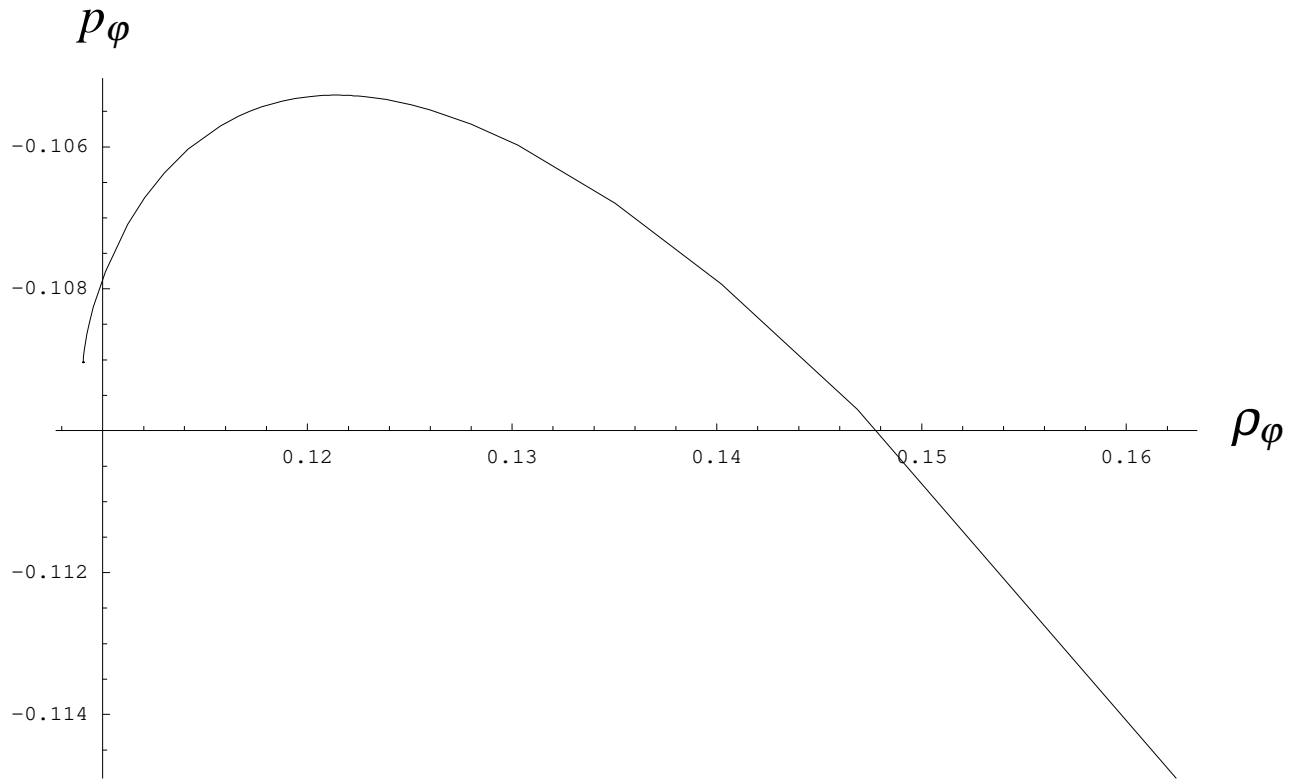
Fig.13 - Case II . State equation for the scalar field.
      Units are arbitrary, but coherent with Fig. 12.

\end{document}